\documentstyle [fleqn,12pt]{article}
\newcommand{\dy}[2]{\frac{{\textstyle \partial \/} #1}{{\textstyle 
\partial \/} #2}}

\begin{document}
\baselineskip .75cm 
\begin{titlepage}
\title{ \bf Strongly Coupled Quark Gluon Plasma (SCQGP) }      
\author{Vishnu M. Bannur \\
{\it Department of Physics}, \\  
{\it University of Calicut, Kerala-673 635, India.} }
\maketitle
\begin{abstract} 
 
We propose that the reason for the non-ideal behavior seen in lattice 
simulation of quark gluon plasma (QGP) and relativistic heavy ion collisions
(URHICs) experiments is that the QGP near $T_c$ and above is strongly 
coupled plasma (SCP), i.e., strongly coupled quark gluon plasma (SCQGP). 
It is remarkable that the widely used equation of state (EoS) of SCP in 
QED (quantum electrodynamics) very nicely fits 
lattice results on all QGP systems, with proper 
modifications to include color degrees of freedom and running coupling 
constant. Results on pressure in pure gauge, 2-flavors and 3-flavors QGP 
are all can be explained by treating QGP as SCQGP as demonstrated here.   
Energy density and speed of sound are also presented for all 3 systems. 
We further extend the model to systems with finite quark mass and a 
reasonably good fit to lattice results are obtained for (2+1)-flavors and 
4-flavors QGP. Hence it is the first unified model, namely SCQGP, 
to explain the non-ideal QGP seen in lattice simulations with just 
two system dependent parameters.       
\end{abstract}
\vspace{1cm}

\noindent
{\bf PACS Nos :} 12.38.Mh, 12.38.Gc, 05.70.Ce, 25.75.+r, 52.25.Kn \\
{\bf Keywords :} Equation of state, quark gluon plasma, 
strongly coupled plasma 
\end{titlepage}
\section{Introduction :}
 
It is now believed that hadrons are confined state of quarks. Fundamental 
theory to explain hadrons is quantum chromodynamics (QCD). 
Non-Abelian nature of QCD leads to complicated non-perturbative 
structure of QCD vaccum. Non-perturbative QCD vaccum leads to 
confinement of quarks at low energy. QCD on lattice confirms this and 
at the same time predicts that, at high energy density or 
baryon density, hadron goes to 
deconfined state called QGP \cite{ga.1}. 
For the last 20 years, lot of experimental, 
theoretical and lattice simulation of QCD results were out, 
but there were no conclusive evidence for QGP and order of phase transition. 
This is probably because the exact nature of matter found in the URHICs   
or in lattice simulations near $T_c$ is not well understood \cite{qgp.1}. 
Both experimental results \cite{gu.1} and lattice results \cite{ka.1} 
show that matter 
formed near $T_c$ is non-ideal even at $T\;>\; T_c$ or above \cite{qgp.1}. 
There are 
lot of attempts to explain such a matter using various models like 
bag model, other confinement models, quasi-particle models, 
strongly interacting quark gluon plasma (sQGP) etc. 

In bag model \cite{rh.1}, QGP is considered 
as a big hadron with large number of 
partons interacting weakly confined by bag wall. This model is only 
partially successful. Inclusions of glue balls or hadrons improve
the results. The confinement models are the extension of bag model with 
smooth potential like Cornel potential \cite{ba.1}, 
relativistic harmonic oscillator \cite{kk.1}  
etc. Again only partially successful. 
In quasi-particle models \cite{pe.1} a new 
concept of calculating thermodynamics of QGP with partons having 
already thermal masses. There are different versions of 
quasiparticle model such as (a) with constant parton masses ($m_G$ or $m_q$) 
and bag constant, $B$ \cite{sk.1},  
(b) with temperature dependent parton masses ($m_G (T)$ or $m_q (T)$) 
and bag constant, $B (T)$, \cite{lh.1,pe.2},  
(c) with further additional function called effective 
degrees of freedon, $D(T)$ or $\nu_g (T)$  to take account of the changes in 
degrees of freedom near $T_c$ \cite{lh.1,s.1}, etc.   
All of them claim to explain lattice results, either by adjusting free 
parameters in the model or by taking lattice data on one of the 
thermodynamic quantity as an input and 
predicting other thermodynamic quantities. However, physical 
picture of quaiparticle model and the origin of various temperature 
dependent quanties are not clear yet \cite{rh.2}.   
In sQGP \cite{sh.1}, 
one considers all possible hadrons even at $T\,>\,T_c$ and try to explain 
non-ideal behavior of QGP near $T_c$. 

Here we propose that the QGP near $T_c$ is in fact what is called 
strongly coupled plasma \cite{ba.2}, 
widely studied in QED plasma. By definition, plasma 
is a quasineutral gas of charged and neutral particles which exhibits 
collective behavior. At sufficiently high temperature neutral particles 
will be negligibly small  so that one can see collective effects of plasma. 
Otherwise it will be just ordinary neutral gas and not plasma.  
SCP \cite{ic.1} is a plasma where the plasma parameter, $\Gamma$,  
defined as the ratio of average potential energy to average kinetic 
energy of the particles, is of the order of 1 or larger.             
Similarly, QGP is a quasi-color-neutral gas of colored particles like 
quarks and gluons which exhibits collective behavior. Here also to see 
collective effects color neutral objects like hadrons and glue balls 
must be negligibly small in number. Otherwise, it is just a hadron gas 
and not QGP. In QED plasma, recombination and ionization are taking 
place with the same rate. Similarly, in QGP such a process may be less 
because QGP is a deconfinement state where confinement effect 
due to QCD vacuum may be melted. Atom is a bound state, 
but hadron is a confinement state. Of course, there will be color 
Coulombic interactions due to one gluon exchange with proper 
quantum effects like running coupling constant etc. and may just 
modify the properties of QGP like EoS etc. In QED plasma 
also particles are interacting via Coulombic force and it modifies EoS.  
Just like in QED plasma, here also 
we can have SCP or SCQGP whenever $\Gamma$ of QGP is of the order of 
1 or greater. Interesting thing is here also interactions potential 
is Coulombic with proper modifications to take account of color degrees 
of freedom and quantum effects. Therefore, 
EoS of QGP may not much different from that of QED plasma. Only 
difference is in $\Gamma$. SCP is widely studied by 
theoretical and numerical methods and EoS of SCP is parametrized 
as function of $\Gamma$ \cite{ic.1}. We will show here that the same EoS with 
$\Gamma$ for QGP, very nicely fits the lattice results on 
EoS of gluon \cite{ka.1}, 
2-flavor as well as 3-flavor QGP \cite{ka.2}. 
Plasma parameters, $\Gamma$, are different for three different 
systems. Further extension of the model to include finite mass of 
the quarks, again, very nicely fits the lattice data on (2+1)-flavors 
and 4-flavors QGP.   
 
\section{Phenomenological Model:}

In Ref. \cite{ba.1}, we attempted to explain the non-ideal behaviour 
seen in lattice by considering Cornel potential, 
Coulomb $+$ linear confining potential, 
using Mayer's cluster expansion method. 
It is only partially successful and near $T_c$, the fit to lattice data 
is not good. This is because Mayer's cluster expansion method is for 
weakly coupled system and hence concluded that QGP near $T_c$ might be 
strongly coupled. In Ref. \cite{ba.2}, we treated QGP as a strongly 
coupled Color-Coulombic plasma and modified EoS of SCP for 
gluon plasma and obtained a remarkably good fit to lattice data. 
Hence we confirmed that QGP near and above $T_c$ is SCQGP and 
speculated that it might have dramatic effects in URHICs observables. Here 
we reconfirm it by fitting lattice datas for all systems, namely, 
gluon plasma, 2-flavor and 3-flavor QGP sucessfully. 

Let us briefly discuss the strongly coupled plasma. An extensive study of 
SCP \cite{ic.1}, theoretical 
and partially numerical, obtained an expression for EoS of SCP as 
a function of $\Gamma$ 
and is given by,
\begin{equation} \varepsilon_{QED} = (3/2 + u_{ex} (\Gamma) ) \, n \, T 
\; , \label{eq:scp} \end{equation}
where the nonideal (or excess) contribution to EoS, 
$u_{ex} (\Gamma)$, is given by, 
\begin{equation} u_{ex} (\Gamma) = \frac{u_{ex}^{Abe} (\Gamma) + 3 \times 10^3 \, \Gamma^{5.7}  
  u_{ex}^{OCP} (\Gamma) }{1 + 3 \times 10^3 \, \Gamma^{5.7} } \; . 
  \label{eq:uex} \end{equation}
Further $u_{ex}^{Abe}$ and $u_{ex}^{OCP}$ are given by   
\begin{equation} u_{ex}^{Abe} (\Gamma) = - \frac{\sqrt{3}}{2} \, \Gamma^{3/2} - 3 \, \Gamma^3 
 \left[ \frac{3}{8} \, \ln (3 \Gamma) + \frac{\gamma}{2} - \frac{1}{3} \right] \; , \end{equation} 
\begin{equation} u_{ex}^{OCP} = - 0.898004 \, \Gamma + 0.96786 \, \Gamma^{1/4} 
      + 0.220703 \, \Gamma^{- 1/4} - 0.86097 \; .  \label{eq:uo} 
      \end{equation}
$u_{ex}^{Abe}$ was derived by Abe \cite{ab.1} exactly in the giant 
cluster-expansion theory and is valid for $\Gamma < .1$. 
$\gamma = 0.57721...$ is Euler's constant.   
$u_{ex}^{OCP}$ was evaluated by computer simulation of one 
component plasma (OCP), where a single species of charged particles 
embedded in a uniform background of neutralizing charges and is valid 
for $1\le \Gamma < 180$. $u_{ex} (\Gamma)$ is valid for all $\Gamma < 180$, 
including the range $.1 \le \Gamma \le 1$. In fact, the final modified 
version \cite{ic.2} of the original numerical 
solutions of hypernetted chain equation (HNC) of Springer {\it et. al.}, 
valid for $.05 \le \Gamma \le 50$ \cite{sp.1}, agrees well with 
$u_{ex}^{OCP}$ with discrepancies of less than $1 \%$.     
$u_{ex} ( \Gamma )$ is derived for strongly coupled Coulombic plasma 
and rigorously verified. 
It may be valid for any Coulombic plasma with appropriate  
change in charge or the coupling constant $\alpha$. 

Let us now consider our model, SCQGP. 
We assume that hadron exists for $T\,<\,T_c$ and goes to QGP for 
$T\,>\,T_c$. That is, for $T\,>\,T_c$ it is the plasma of quarks and 
gluons and no hadrons or glue balls. But it is a strongly coupled plasma, 
which we call SCQGP. The plasma parameter $\Gamma$ is defined as,   
\begin{equation} \Gamma \equiv \frac{<PE>}{<KE>} = \frac{ \frac{4}{3} 
\frac{\alpha_s}{r_{av}} }{T} \; ,\end{equation}
where we have taken Coulombic interaction between quarks, generally used in 
hadron spectroscopy. The typical value of $\alpha_s \approx 0.5$, 
$r_{av} \approx 1 fm$ and near the critical temperature, $T_c \approx 
200 \; MeV$, we estimate $\Gamma \approx 2/3$. Hence QGP is a strongly 
coupled plasma. Later we will see that the $\Gamma$ 
which fits lattice results is indeed of the order of 1. 
Compared to QED, the fine structure constant, $\alpha$, 
is replaced by $4 \; \alpha_s /3$ in Coulombic interaction term. 
$r_{av}$ may be estimated as $r_{av} = (3/4 \pi n)^{1/3}$   
and hence  
\begin{equation} \Gamma = \left( \frac{4 \pi n}{3} \right) ^{1/3} \frac{4}{3} \frac{\alpha_s}{T} 
\; , \label{eq:gaa}  \end{equation}
where '$n$' is the number density. 

In SCP, where one generally has high enough temperature, electrons are 
in continuum state of atoms, i.e, ionized state, and negligibly small 
amount of neutral atoms. In SCQGP also there may be negligibly small 
hadrons due to Coulomb binding interactions. Since it is a deconfined 
state, there is no confinement interactions. Note the difference 
between sQGP and SCQGP. In sQGP presence of various neutral 
and colored bound states due to Coulomb interactions give rise to 
non-ideal effects.  

Our phenomenological model for SCQGP is obtained by modifying 
Eq. (\ref{eq:scp}), 
to include relativistic and quantum effects as, 
\begin{equation} \varepsilon = (2.7 + u_{ex} (\Gamma) ) 
\, n \, T \;, \end{equation}
where the first term $2.7 \,n\, T$ corresponds to the ideal EOS which, in 
our case, may be written as $\varepsilon_s \equiv 3 a_f T^4$, 
ideal EoS for massless relativistic gas. 
$a_f \equiv (16 + 21 \, n_f /2) \pi ^2 /90$ is a constant 
which depends on degrees of freedom. Hence we write,  
\begin{equation} e(\Gamma) \equiv \frac{\varepsilon}{\varepsilon_s} 
= 1 + \frac{1}{2.7} u_{ex} (\Gamma) \; . \label{eq:e} \end{equation} 
We assume that the functional form of $u_{ex} (\Gamma)$ is the same as 
that of SCP, but $\Gamma$, which follows from Eq. (\ref{eq:gaa}), is, 
\begin{equation} \Gamma \equiv \left( \frac{4.4 \,\pi a_f}{3} 
\right) ^{1/3} g_c \alpha_s (T) 
\; ,  \label{eq:gab}  \end{equation}
where we have taken $n \approx 1.1 \,a_f T^3$. 
This is motivated by the fact that SCQGP is also governed by Coulombic 
type interactions, i.e, Color Coulombic interaction. Only new thing is 
$g_c$ and $\alpha_s (T)$. In SCP $g_c$ is 1 and in SCQGP it is different 
for gluon plasma and flavored QGP because of different group 
structures involved in their interactions. Similarly, 
$\alpha_s = 1/137 =$constant in SCP, but in SCQGP, including quantum  
effects, it is a running coupling constant $\alpha_s (T)$, given by, 
\begin{equation} \alpha_s (T) = \frac{6 \pi}{(33-2 n_f) \ln (T/\Lambda_T)}  
\left( 1 - \frac{3 (153 - 19 n_f)}{(33 - 2 n_f)^2} 
\frac{\ln (2 \ln (T/\Lambda_T))}{\ln (T/\Lambda_T)} 
\right)  \;, \end{equation}
where $n_f$ is the number of flavors and $\Lambda_T$ is a parameter,  
related to QCD scale parameter. Since lattice results are with 
two-loop order running coupling constant, we have chosen similar form.  

Finally, we have a phenomenological EoS for SCQGP with two parameters 
$g_c$ and $\Lambda_T$. Both of them depends on type of QGP. Any 
shortcoming may be reflected in the correctness of these parameters. 
From Eq. (\ref{eq:e}), we get 
$\varepsilon (T)$ and using the relation, 
$\varepsilon = T \dy{\textstyle P}{\textstyle T} - P \; ,$ 
we get the pressure 
\begin{equation} \frac{P}{T^4} = \left( \frac{P_0}{T_0} + 3 a_f \int_{T_0}^T \, 
d\tau \tau^2 e(\Gamma(\tau)) \right) / T^3 \; , \label{eq:p} \end{equation} 
where $P_0$ is the pressure at some reference temperature $T_0$.
and may be fixed 
to one of the lattice data points or at critical temperature $T_c$. 
Once we know $P$ and 
$\varepsilon$, $c_s^2 = \dy{\textstyle P}{\textstyle \varepsilon}$ 
can be evaluated. 

In the case of QGP with finite quark masses, the $n_f$ in the expression for 
$a_f$ and hence in $\Gamma$ must be replaced by effective $n_f$, $n_f^{eff}$ as 
discussed in \cite{ka.2,ka.3}. 
   
\section{Results :} In Fig. 1, we plotted $P(T) / T^4$ Vs $T$ for 
pure gauge, 2-flavor and 3-flavor QGP along with lattice results. 
Note that, in the case of flavored QGP, since there is 
$(10\% \pm 5\%)$ uncertainty in $P$ data \cite{ka.3} on taking continum limit 
with massless quarks \cite{ka.3}, we multiply the lattice 
data by the factor $1.1$ and is plotted. For each system $g_c$ and 
$\Lambda_T$ are adjusted so that we get a good fit to lattice results. 
We have fixed $P_0$ from the lattice data at the critical temperature $T_c$ 
for each system. Surprisingly good fit is obtained for 
all systems with $g_c = 1.4$ 
for gluon plasma and $g_c = 0.89$ for both 2-flavor and 3-flavor QGP. 
$\Lambda_T$ are different for all systems with values 137.5, 80.5 
and 37 for gluon plasma, 2-flavor and 3-flavor QGP respectively. 
We have taken $n_f$ equal to $0$, $2$ and $3$ respectively for three systems. 
The values of $g_c$ are not that unreasonable since they are close to 
the eigen values of quadratic Casimir operators, 
$3$ for pure gauge and $4/3$ for quarks.  

Once $P(T)$ is obtained, then other macroscopic quantities such as 
$\varepsilon$, $c_s^2$ etc.  are derivable from $P(T)$ and no other 
parameters are needed. In Fig. 2, we plotted $\varepsilon / T^4$ 
Vs $T/ T_c$ for all three systems along with lattice 
results and a resonably good fit is obtained without any extra parameters. 
All the three curves looks similar, but shifts to left 
as flavor content increases. We have taken $T_c$ equal to $275$, $175$ and 
$155$ $MeV$ respectively for gluon plasma, 2-flavor and 3-flavor QGP.  

In Fig. 3,  $c_s^2$ is plotted for all three systems, again with 
lattice results for gluon plasma. Reasonably good fit for gluon plasma 
and our predictions for the flavored QGP. All the three curves have 
similar behaviour, i.e, sharp rise near $T_c$ and then flatten to the 
value close to $1/3$. $c_s^2$ is larger for larger flavor content.  
Very close to $T=T_c$,  
fits or predictions of our model may not be good, especially for 
$\varepsilon$ and $c_s^2$ which strongly depends on variations of 
$P$ with respect to $T$. Lattice data also has large error bars 
very close to $T_c$. However, except for small region at 
$T=T_c$, our results are very good for all regions of $T > T_c$.

In Fig.s 4 and 5, we plotted $\alpha_s $ and $\Gamma $ respectively 
as a function of $T/T_c$. We see that QGP near $T>T_c$ upto several 
$T_c$ is really strongly coupled since $\Gamma$ is the order of 
1. We see from the plot that pure gauge is more strongly coupled 
than flavored QGP. It is interesting to note that recently  
Peshier and Cassing \cite{pe.3} also obtained similar results on 
$\Gamma$ as a function of $T$ in quai-particle model and concluded that 
QGP behaves like a liquid.    

In Fig.6, we included the results on (2+1)-flavors and 4-flavors QGP 
from our model along with lattice data \cite{ka.3,ka.4} and 
replotted $P(T)/T^4$ Vs $T/T_c$ for all systems. 
Similar plots for $\varepsilon (T)/T^4$ Vs $T/T_c$ for all systems 
is replotted in Fig. 7. The fitted parameters for various systems are 
tabulated in Table 1. The effective number of flavors, $n_f^{eff}$ is   
$(2\,\,\, 0.9672 + 0.8275)$ for (2+1)-flavor QGP 
and $4\,\,\, 0.9672$, $4\,\,\, 0.9915$ for 
4-flavor QGP with mass $m/T$ equal to $0.4$ and $0.2$ respectively. 
Masses $m/T$ in the case of (2+1)-flavors are $0.4$ and $1.0$ 
for light and heavy quarks respectively. 

\section{Conclusions :} 

Using a phenomenological model to treat QGP near and above $T=T_c$ 
as SCQGP, obtained by appropriate modifications of the results of 
SCP to take account of 
color and flavor degrees of freedom and quantum effects,
a surprisingly good fits to lattice results are obtained. Basic idea is that 
SCP and SCQGP both are nonideal because of Coulombic interactions. 
So we expect similar EoS. Modification of SCP to SCQGP introduces 
system dependent two parameters which we vary to get good fit to 
lattice results for pressure $P$. $\varepsilon$, $c_s^2$ etc. then follows 
from $P$ without any extra parameters. When we extend our model to 
include systems with finite quark mass, again, a remarkable good fits 
are obtained to (2+1)-flavors and 4-flavors QGP. Hence we have a unified 
model with just two system dependent parameters, namely SCQGP, 
to explain the non-ideal effects seen in the lattice 
simulation of QCD in various systems like pure gauge, 2-flavors, 3-flavors, 
(2+1)-flavors and 4-flavors QGP .        
Of course, to understand the values 
of the parameters, $g_c$ and $\Lambda_T$, one need much more 
general analytic theory based on QCD. 
It is also interesting to see 
the modifications of URHICs results by treating QGP as SCQGP, since the 
energy density and hence temperature is in the range of few $T_c$. \\[.5cm] 

\noindent 
{\bf Acknowledgement:} 

I thank the organizer and DAE of India for the financial help to attend 
the International conference on ICPAQGP at Kolkata, India, which 
inspired me to present this result.

\newpage
\begin{figure}
\caption { Plots of $P/ T^4 $ as a function of $T$ from 
our model and lattice results (symbols) for pure gauge (lower curve), 
2-flavor QGP (middle curve) and 3-flavor QGP (upper curve). } 
\label{fig 1}
\vspace{.75cm}

\caption { Plots of $\varepsilon/ T^4 $ as a function of $T/T_c$ from 
our model for pure gauge (lower curve), 
2-flavor QGP (middle curve) and 3-flavor QGP (upper curve) and also 
with lattice data for pure gauge. } 
\label{fig 2}
\vspace{.75cm}

\caption { Plots of $c_s^2$ as a function of $T/T_c$ from 
our model for pure gauge (lower curve), 
2-flavor QGP (middle curve) and 3-flavor QGP (upper curve) and also 
with lattice data for pure gauge. } 
\label{fig 3}
\vspace{.75cm}

\caption { Plots of $\alpha_s $ as a function of $T/T_c$ from 
our model for pure gauge (upper curve), 
2-flavor QGP (middle curve) and 3-flavor QGP (lower curve). } 
\label{fig 4}
\vspace{.75cm}

\caption { Plots of $\Gamma $ as a function of $T/T_c$ from 
our model for pure gauge (upper curve), 
2-flavor QGP (middle curve) and 3-flavor QGP (lower curve). } 
\label{fig 5}
\vspace{.75cm}

\caption { Plots of $P/ T^4 $ as a function of $T/T_c$ from 
our model and lattice results (symbols) for pure gauge, 
2-flavor QGP, (2+1)-flavor QGP, 3-flavor QGP and 4-flavor QGP 
(with two different masses). } 
\label{fig 6}
\vspace{.75cm}

\caption { Plots of $\varepsilon / T^4 $ as a function of $T/T_c$ from 
our model and lattice results (symbols) for pure gauge, 
2-flavor QGP, (2+1)-flavor QGP, 3-flavor QGP and 4-flavor QGP 
(with two different masses). } 
\label{fig 7}
\vspace{.75cm}

\end{figure}
\vspace{5cm}
\begin{center}
{\bf Table 1} \\[1cm]
Parameters of our model for various systems to get 
the best fit to lattice data. \\[1cm] 
\begin{tabular}{||c|c|c|c||}
\hline
systems &$t_0 \equiv \Lambda_T /T_c$ &$g_c$ \\
\hline
pure gauge&0.5 &1.4 \\
\hline
2-flavor&0.46 &0.89 \\
\hline
3-flavor&0.24 &0.89 \\
\hline
(2+1)-flavor&0.24 &1.3 \\
\hline
4-flavor($m/T =.4$)&0.55 &0.36 \\
\hline
4-flavor($m/T =.2$)&0.55 &0.44 \\
\hline
\end{tabular}
\end{center}
                                                                                
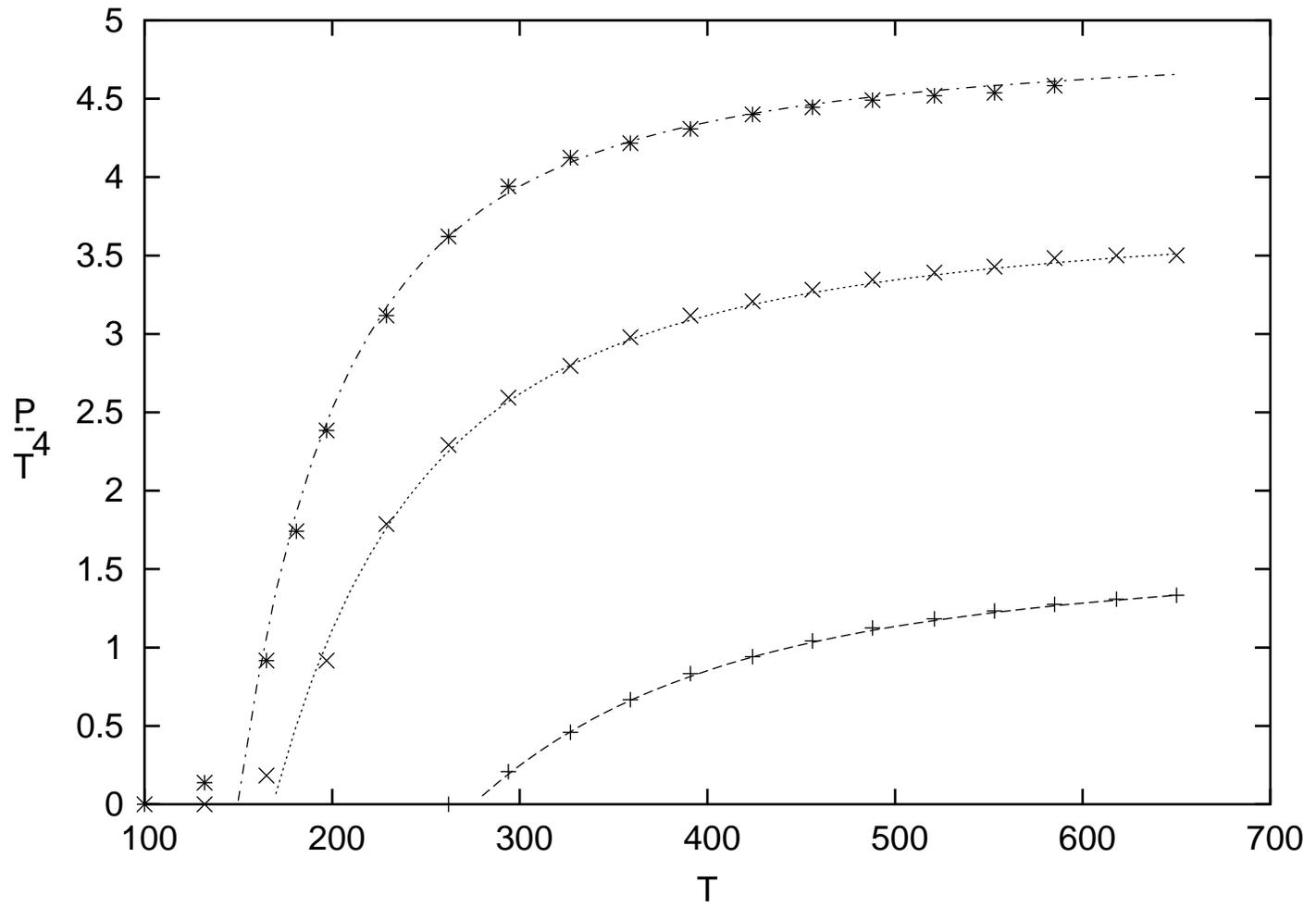
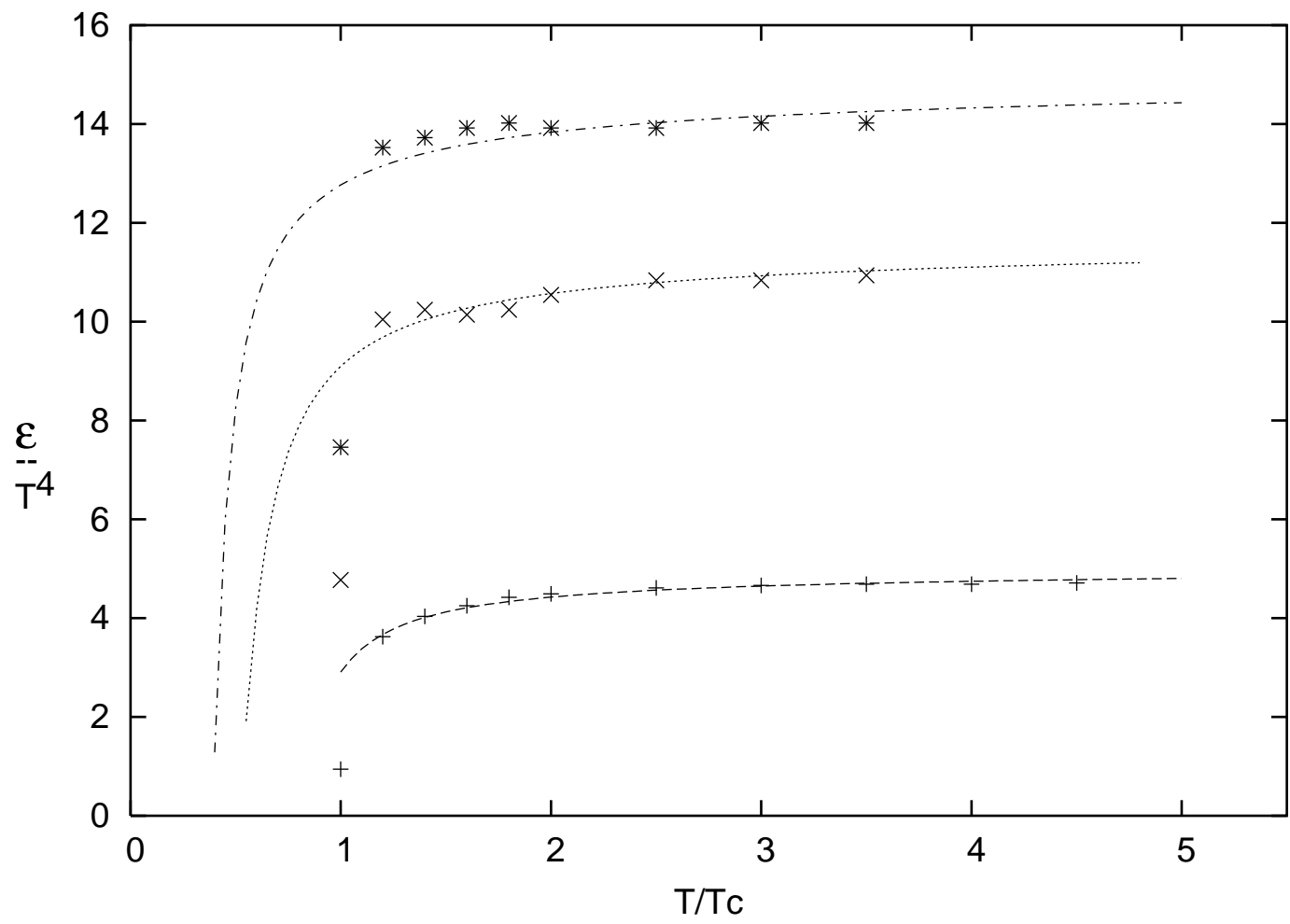
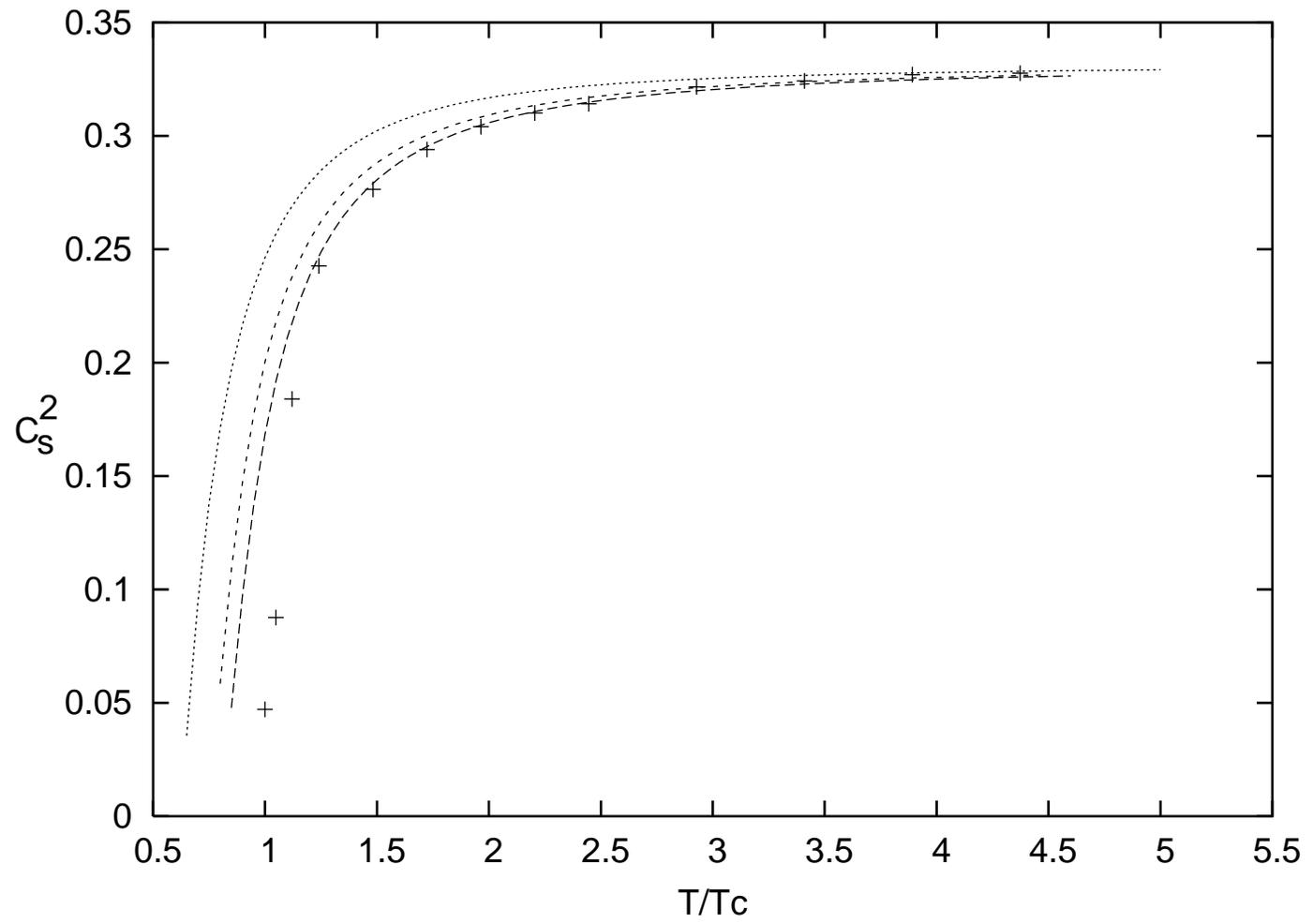
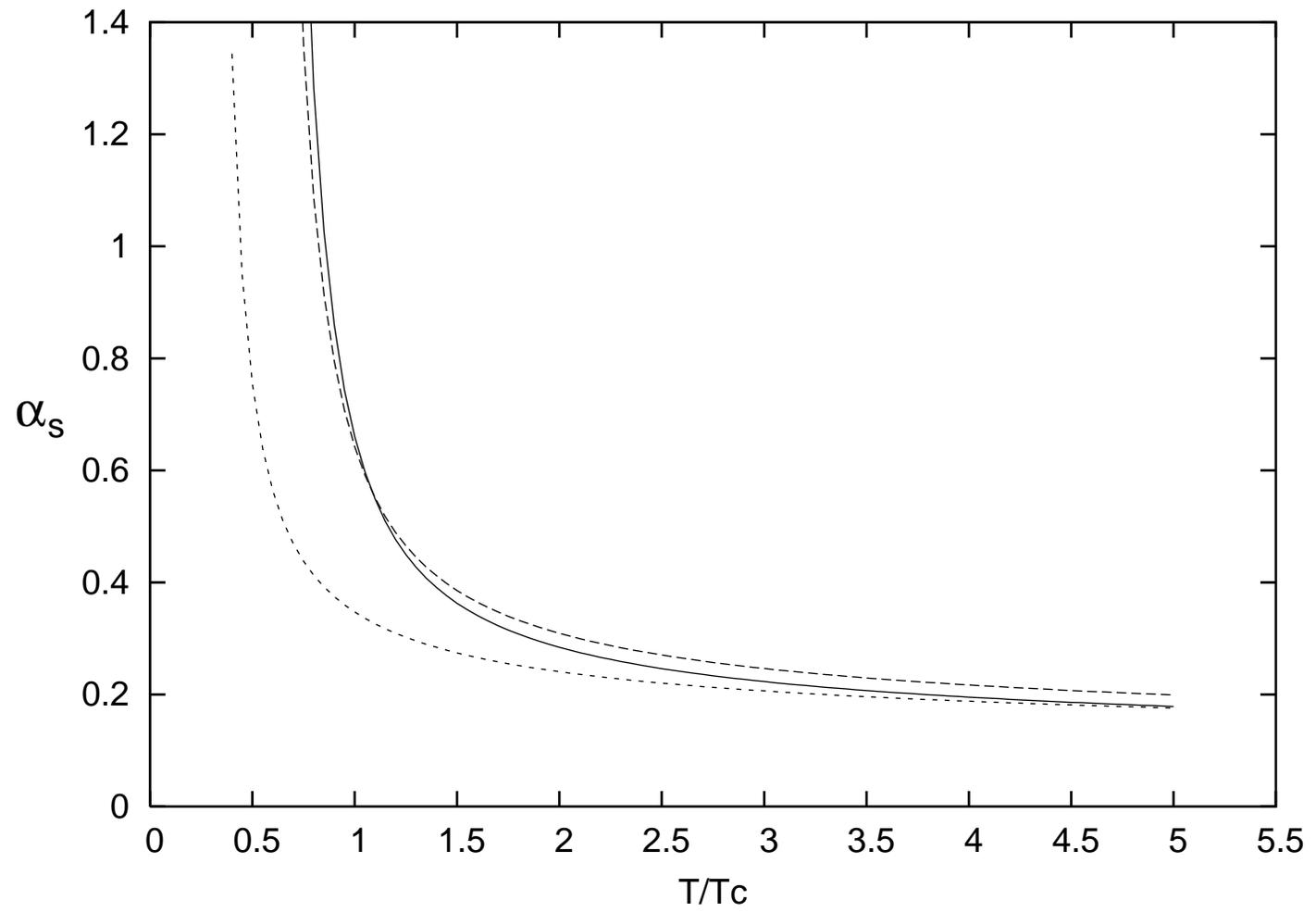
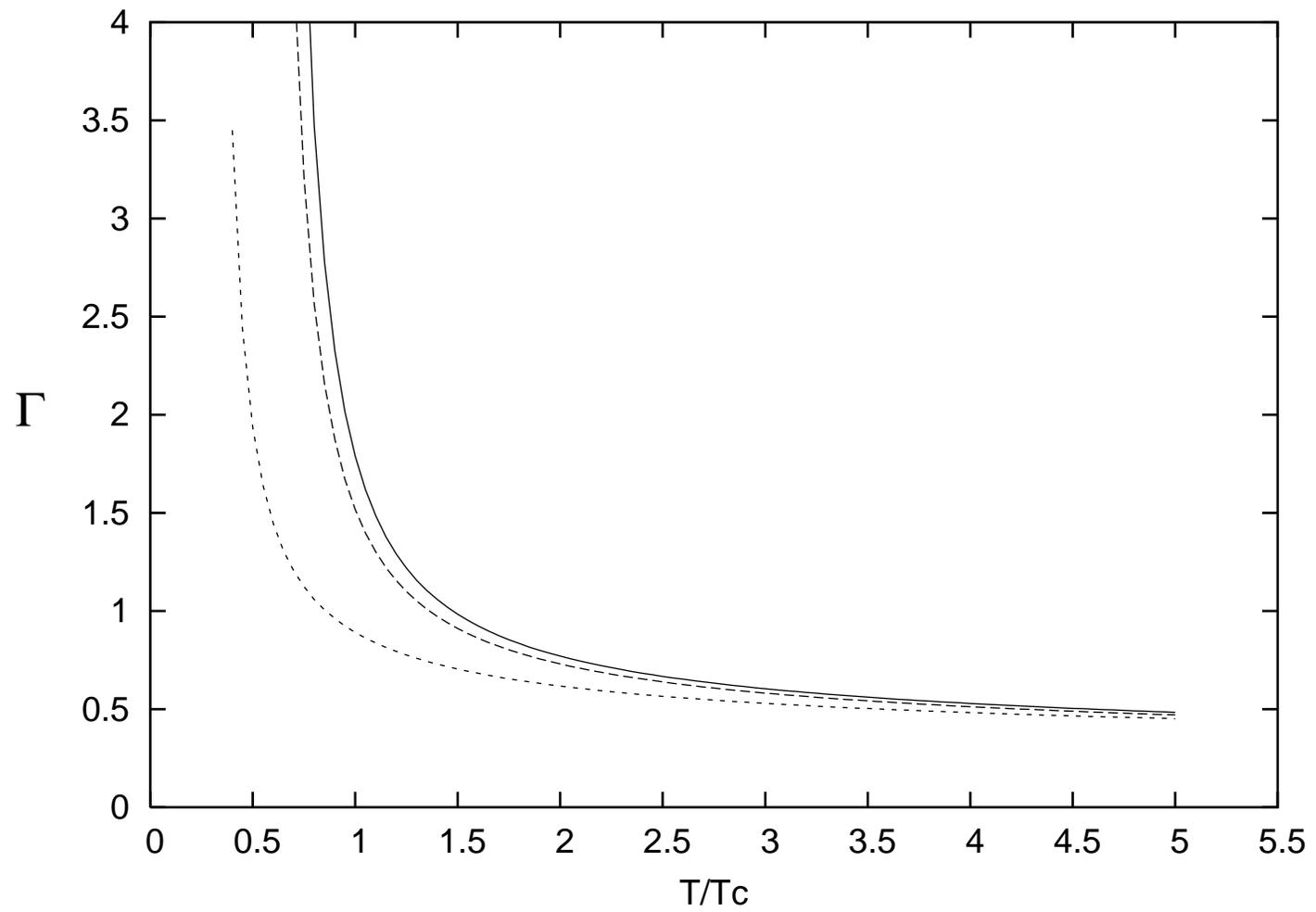
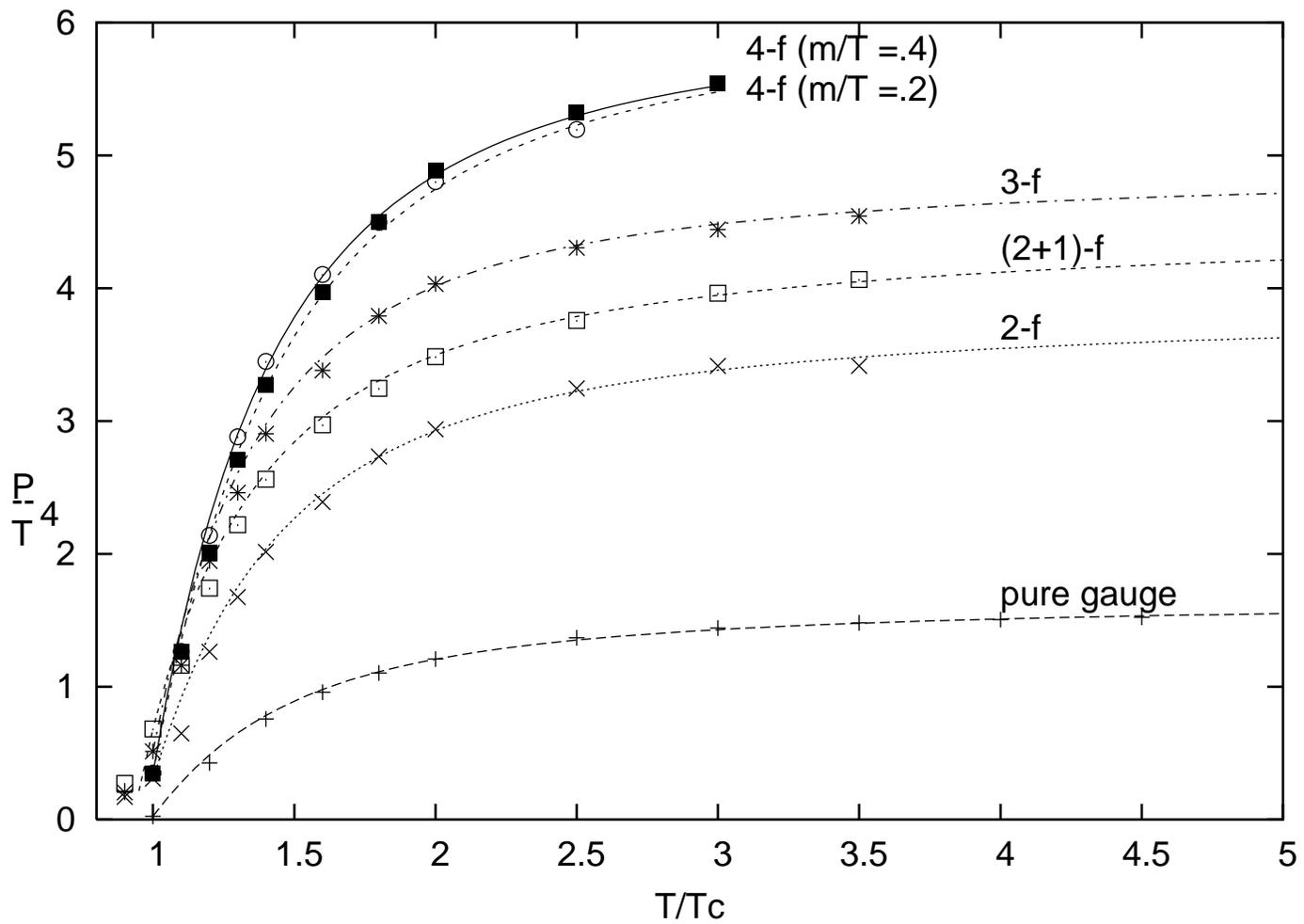
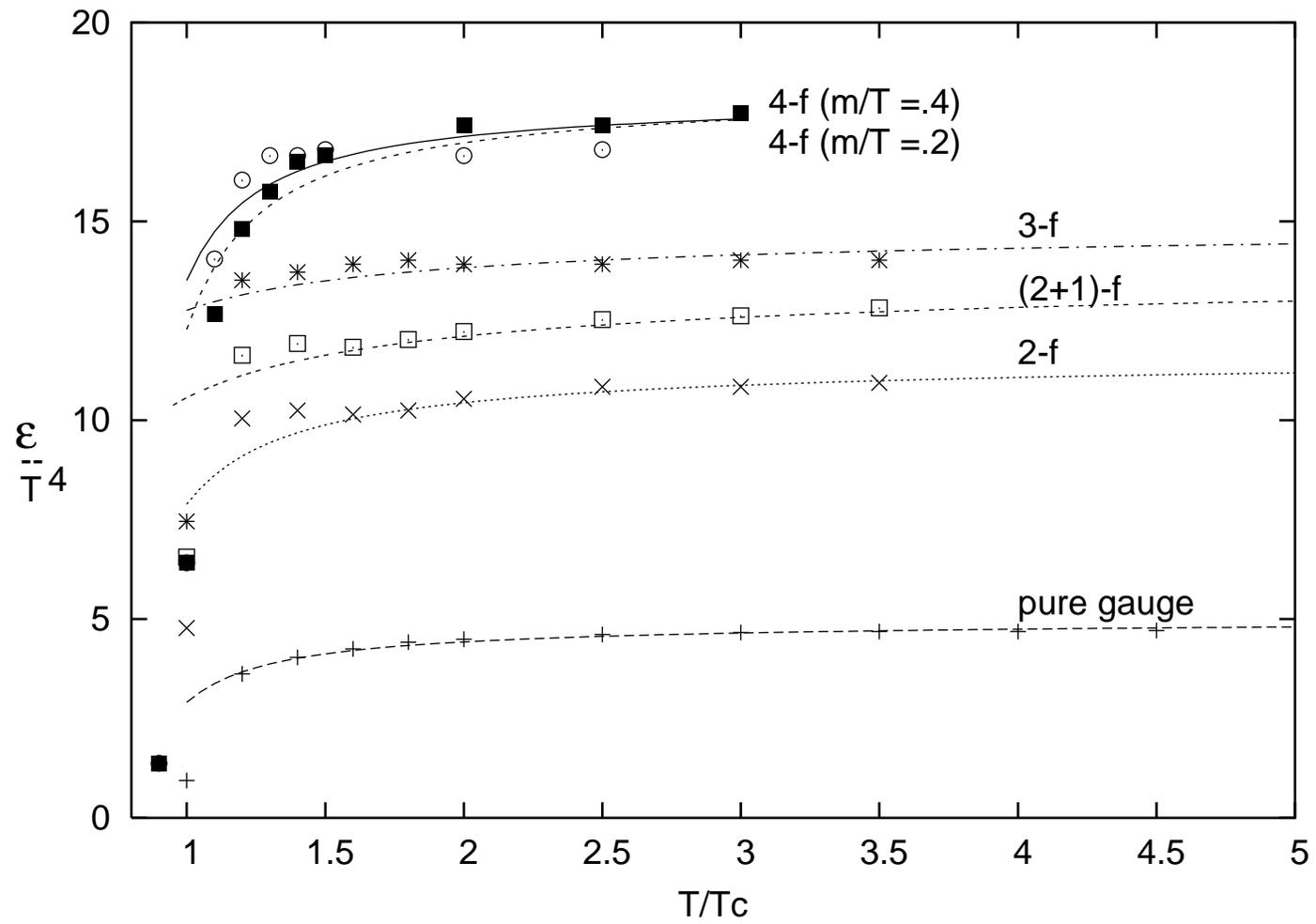
\end{document}